# Adsorption of organic molecules at the $TiO_2$(110) surface: the effect of van der Waals interactions

Marcus Tillotson,[1,2*] Peter Brett[2], Roger A. Bennett[3], and Ricardo Grau-Crespo[3†]

[1]*Department of Chemistry, University College London, London, WC1H 0AJ, United Kingdom.*
[*]*E-mail: marcus.tillotson.09@ucl.ac.uk*

[2]*Biomaterials and Tissue Engineering, UCL Eastman Dental Institute, 256 Gray's Inn Road, London, WC1X 8LD, United Kingdom.*

[3]*Department of Chemistry, University of Reading, Whiteknights, Reading RG6 6AD, United Kingdom.*
[†]*E-mail: r.grau-crespo@reading.ac.uk*

**Abstract**. Understanding the interaction of organic molecules with $TiO_2$ surfaces is important for a wide range of technological applications. While density functional theory (DFT) calculations can provide valuable insight about these interactions, traditional DFT approaches with local exchange-correlation functionals suffer from a poor description of non-bonding van der Waals (vdW) interactions. We examine here the contribution of vdW forces to the interaction of small organic molecules (methane, methanol, formic acid and glycine) with the $TiO_2$ (110) surface, based on DFT calculations with the optB88-vdW functional, which incorporate non-local correlation. The adsorption geometries and energies at different configurations were also obtained in the standard generalized gradient approximation (GGA-PBE) for comparison. We find that the optB88-vdW consistently gives shorter surface adsorbate-to-surface distances and slightly stronger interactions than PBE for the weak (physisorbed) modes of adsorption. In the case of strongly adsorbed (chemisorbed) molecules both functionals give similar results for the adsorption geometries, and also similar values of the relative energies between different chemisorption modes for each molecule. In particular both functionals predict that dissociative adsorption is more favourable than molecular adsorption for methanol, formic acid and glycine, in general agreement with experiment. The dissociation energies obtained from both functionals are also very similar, indicating that vdW interactions do not affect the thermodynamics of surface deprotonation. However, the optB88-vdW always predicts stronger adsorption than PBE. The comparison of the methanol adsorption energies with values obtained from a Redhead analysis of temperature programmed desorption data suggests that optB88-vdW significantly overestimates the adsorption strength, although we warn about the uncertainties involved in such comparisons.

## Introduction

The adsorption of organic molecules at the surface of titanium oxide is an important phenomenon for many areas of modern technology, including photocatalysis [1], dye-sensitised solar cells [2], and biological implants [3]. In the latter field, titanium and its alloys have long been the material of choice for the repair of damaged or defective bone tissue. Close interactions occur between prototype adhesion proteins on bone cell surfaces and the oxide layer at the surface of the metal implant [4]; the important adhesion process is then controlled by the strength of the molecular interactions with the oxide surface. The current understanding of the interactions of organic molecules with $TiO_2$ surfaces has been summarized in a recent review by Thomas and Syres [5].

Rutile $TiO_2$ (110) has been one of the most widely studied surfaces in surface science [6,7]. It is the most stable surface in rutile, and the structure of its bulk-like (1×1) termination is well known from both experiment and theory [6]. This surface termination exhibits alternating rows of twofold coordinated bridging oxygen atoms and channels that expose fivefold coordinated titanium ($Ti_{5f}$) atoms with in-plane threefold coordinated O atoms (Fig. 1). In this work, we use the (110) surface as a representative $TiO_2$ surface, to investigate the role of van der Waals (vdW) forces in the interaction between oxides surfaces and (small) organic molecules. Establishing the correct theoretical framework for studying these interactions is a necessary first step in order to progress to more complex computer simulations of the adhesion of biological molecules at the $TiO_2$ surfaces.

Earlier theoretical studies of the interaction of organic molecules with $TiO_2$ surfaces using density functional theory (DFT) have contributed to understanding the main adsorption modes and provided useful estimations of the relative adsorption strengths at different sites [8-12] but they suffer from the poor description of vdW forces within the traditional DFT framework. Both the local density approximation (LDA) and the generalized gradient approximation (GGA) fail to provide the long-range ($r^{-6}$) attractive dispersion term, which arises from non-local electron-electron correlations. In recent years, several schemes have been proposed to add dispersion corrections to existing DFT formulations [13-18]. Among them, the so-called van der Waals density functional (vdW-DF) method (which is the focus of this paper), is particularly promising because it provides the dispersion correction based directly on the electron density.

The vdW-DF method was originally proposed by Dion et al. [14], and the algorithm was improved by Roman-Perez and Soler to reduce the computational cost [19]. In this functional, the exchange-correlation term is calculated as:

$$E_{\mathrm{xc}} = E_{\mathrm{x}}^{\mathrm{GGA}} + E_{\mathrm{c}}^{\mathrm{LDA}} + E_{\mathrm{c}}^{\mathrm{nl}} \quad (2)$$

where $\mathrm{E}_{\mathrm{x}}^{\mathrm{GGA}}$ is the GGA exchange energy and $E_{\mathrm{c}}^{\mathrm{LDA}}$ is the local correlation energy obtained with the local density approximation (LDA). $\mathrm{E}_{\mathrm{c}}^{\mathrm{nl}}$ is the non-local correlation energy based on electron densities interacting via a model response function, and is the key ingredient for the description of dispersion interactions. In the original vdW-DF, the GGA exchange energy was given by the revPBE functional [20] (a modified version of the standard GGA functional of Perdew, Burke and Ernzerhof, PBE [21]). Klimeš et al. later showed that the choice of exchange functional has an important effect on the interaction energies given by the vdW-DF [22]. Therefore they tried a series of different exchange functionals, keeping the non-local correlation as in the original vdW-DF, to optimize the agreement with reference data (interaction energies in a set of weakly interacting dimers and water clusters, obtained from coupled cluster calculations). The opt-B88-vdW functional, which we employ in the present study, is one of these optimized functionals, and it uses the exchange part of the B88 functional [23]. It was found that this functional provided much better overall accuracy than the original vdW-DF (and of course, than standard GGA approaches) in the description of weakly interacting dimers in the gas phase [22]. A later study showed that the optB88-vdW functional also provided good values for the lattice constants and atomization energies of “hard” solids, performing better for this type of calculations than the original vdW-DF, and similar or slightly better than PBE [24].

Capitalising on this recent progress, there have been a number of studies examining the effect of dispersion on surface-adsorbate interactions at oxides. For example, Plata *et al.* found that the adsorption energies of acetaldehyde molecules on the $TiO_2$ solid surface were significantly larger when computed with the optB86b-vdW functional (another one of the modified vdW functionals optimized by Klimeš et al. [22]) than when obtained with local GGA functionals [25]. In addition, they found that the adsorption energy of individual acetaldehyde molecules decreases with increasing surface coverage, due to lateral interactions between neighbouring adsorbed species. These adsorbate-adsorbate lateral interactions are mainly attractive dispersion forces; thus, calculations involving the optB86b-vdW functional produced a significantly stronger adsorption at high coverage compared to the GGA functional, and in better agreement with experiment. Sorescu et al. investigated the adsorption of $CO_2$ on $TiO_2$(110) and found that adsorption was predicted to be stronger when calculated with several dispersion-corrected functionals as compared with GGA calculations, bringing the theoretical results into closer agreement with experiment [26]. In another recent work by Johnston [27] various vdW functionals were employed to calculate the adsorption energy and geometries of ethanol at the α-$Al_2O_3$ (0001) surface. This study also found that vdW functionals predicted stronger adsorption than the GGA functional, although both methods predicted dissociative adsorption (with transfer of a proton to the surface) to be more stable than molecular adsorption.

It is clear from the results of these studies that DFT calculations with vdW functionals can lead to significantly different energies and geometries of adsorbed molecules at oxide surfaces, compared to standard GGA calculations. The expectation, based on the tests mentioned above, is that the results from the vdW functionals are more reliable than those from the GGA functionals, at least in the case of physisorbed and weakly chemisorbed molecules. However, the direct benchmarking of the performance of different functionals against experimental adsorption reference data is generally difficult due to uncertainties in the interpretation of experimental measurements, as has been shown for the case of benzene adsorption on Cu(111) [28]. In an encouraging success story, a recent study has shown that the optB88-vdW functional provides excellent agreement with the experimentally derived binding energy for a polycyclic aromatic hydrocarbon molecule on a graphitic surface [29]. The comparison required a sophisticated interpretation of the temperature programmed desorption (TPD) data, taking into account entropic effects, to derive realistic adsorption energies.

In the present paper, our aim is to investigate the effect of vdW forces on the adsorption of small organic molecules at the $TiO_2$(110) surface. In order to focus on the direct adsorbate-surface interactions, rather than the lateral interactions, our calculations are undertaken at low coverage. The paper is organised as follows: we first discuss the equilibrium relaxed structures of $TiO_2$(110) in comparison with experimental data from low energy electron diffraction (LEED) and surface x-ray diffraction (SXRD). We then discuss the energies and geometries for different adsorption configurations of a number of small organic molecules containing alkyl, hydroxyl, carboxyl and amino groups, as obtained from the simulations with normal GGA calculations and with a vdW density functional. We compare our results with those from previous studies, and, whenever possible, with experimental results.

# 1. Methodology

## *1.1 Structural models*

The oxide surface was simulated using a periodic stack of quasi-two dimensional slabs, where each slab is separated from the neighbour by a vacuum gap. The gap width was set to 15 Å to provide sufficient space for the adsorbates, avoiding their interactions with neighbouring (periodic image) slabs. We used slabs with four O-Ti-O tri-layers (figure 1), as in several previous theoretical studies of this surface [11,30-32]. Some authors have pointed out that calculated surface energies display significant oscillations with the number of layers and are not well converged for this slab thickness. For example, Kiejna et al. reported an increase of 39% in the surface energy when increasing the slab thickness from 4 to 5 trilayers [33]. The origin of this behaviour is in the relative positions of the ions at each side of the slab, which is different for even and odd numbers of layers. In even-layer slabs the five-fold Ti atom in the top surface is directly above the six-fold Ti atom in the bottom surface, while in odd-layer slabs equivalent Ti atoms at the top and bottom surfaces are exactly opposite each other. Therefore, in even-layer slabs the displacements of Ti atoms at each side of the slab reinforce each other, whereas in odd-layer slabs they oppose each other [34]. In order to counteract this effect and avoid interactions between the relaxations of the two surfaces, we kept the bottom two layers of our slabs fixed to their bulk positions, while the ions in the top two layers were allowed to relax to their equilibrium positions, as illustrated in figure 1. This setting has been shown to increase the convergence of calculated properties on $TiO_2$(110) with respect to the number of layers in the simulation slab [35]. We found that the increase in the GGA surface energy from 4- to 5-layer calculations was of 13% (from 0.65 to 0.73 $Jm^{-2}$), much less than the increase reported by Kiejna et al. Moreover, we have tested the convergence of the adsorption energies with respect to slab thickness in the case of methanol on different binding modes, and with both functionals, and found that the addition of an extra layer in the slab has a very small effect on our results (see section 3.3 below).

In the directions parallel to the surface, the simulation cell contained 4×2 surface unit cells, which means that the distance from the adsorbed molecule to its periodic images is at least ~12 Å, and therefore lateral interactions should be small. The adsorption of one molecule at this surface cell corresponds to 0.125 monolayer (ML) coverage. Adsorption energies $E_{ads}$ were then found using the equation:

$$E_{ads} = E_{slab\ +\ molecule} - E_{slab\ +\ molecule\ far} \tag{1}$$

where $E_{slab\ +\ molecule}$ is the energy of the molecule / surface complex and $E_{slab\ +\ molecule\ far}$ is the energy of the same system with the molecule placed in the centre of the vacuum gap, where the interaction with the surface is very small. This approach allows an effective cancellation of errors from the two terms in the right-hand side of equation (1), and also cancels out the small contributions from lateral interactions between the molecule images to $E_{ads}$. Thus, our results refer to the limit of low coverage.

*2.2 Density functional theory calculations*

All the calculations were performed using the plane-wave DFT code VASP [36,37]. Geometries and total energies were obtained both with the generalised gradient approximation (GGA) with the exchange-correlation functional of Perdew, Burke and Ernzerhof (PBE) [21], and with the opt-B88-vdW functional, which provides a good description of van der Waals interactions [22].

Formally the Ti cations in the $TiO_2$ bulk and (110) surface are in the 4+ oxidation state, where the 3*d* orbitals are empty. In practice, due to covalent effects and, in the case of the surface simulations, to small charge transfer from adsorbates, the occupancy of the Ti 3*d* levels is not zero. We therefore applied a Hubbard-type correction to the orbitals, both in the PBE and in the optB88-vdW calculations. This correction, which is proportional to a parameter $U_{eff}$, has the effect of penalising the hybridisation of the specified orbitals of the metal atoms (Ti 3*d* orbitals) with the ligands (O atoms), thus counteracting the artificial delocalisation that results from the spurious electron self-interaction in local-exchange DFT [38]. A value of $U_{eff}$ = 3 eV was used in this work, as in some previous studies [39-41]. This value was found by Nolan et al. to be optimal to reproduce the experimental electronic structure of Ti rich nonstoichiometric $TiO_2$ surfaces with partial 3*d* occupancy [39]. Other values of $U_{eff}$ in GGA-based functionals have been suggested by other authors (e.g. 2.5 eV [42], 3.5 eV [43] and 4.2 eV [44]), but for the systems considered in the present work, where all the Ti atoms formally remain in the 4+ oxidation state, the change in our results due small variations of $U_{eff}$ can be expected to be negligible.

The interaction between the valence electrons and the core was described with the projected augmented wave (PAW) method in the implementation of Kresse & Joubert [45]. The core electrons, up to 3p in Ti and 1s in O were kept frozen in their atomic reference states. The cut-off energy controlling the number of plane wave basis functions was set to $E_{\text{cut}} = 400$ eV. Due to the size of the simulation cell (shortest cell vector is ~12 Å), reciprocal space integration during ion relaxations were performed by sampling only the Γ point of the Brillouin zone. Test calculations confirmed that including more k-points in the integration led to negligible or zero effect on the relaxed geometries. Final energies were then obtained with a single-point calculation and a denser k-point mesh of 3×3×1 divisions. Relaxations of the ion positions were performed using a conjugate gradient algorithm, until the forces on all atoms were less than 0.01 eV/ Å. A dipole correction was applied in the direction perpendicular to the surface.

## 2. Results and discussion

### *3.1 Relaxations in the clean surface*

We first compare the calculated atomic displacements due to surface relaxations, in the direction perpendicular to the surface, with experimental results. In Table 1 we list the theoretical displacements obtained with both the PBE functional and the optB88-vdW, in comparison with data from quantitative LEED [46] and SXRD [37] experiments. There is good agreement between the two sets of experimental results, with only small differences in the relaxation shift of the $Ti_{5f}$ and the in-plane O atoms.

From our calculations we find that the displacements of the surface atoms are well described by both the PBE and optB88-vdW approximations for the exchange and correlation functionals, leading to good agreement with the experimental work. The optB88-vdW functional predicts slightly shorter outward relaxations of the $Ti_{6f}$ and the in-plane O layers compared to the PBE functional, which gives values closer to experiment in both cases. The inward displacement of the $Ti_{5f}$ atom is predicted by PBE to be in good agreement with the SXRD determination, and for the optB88-vdW it was intermediate between the SXRD and the LEED value. For the $O_b$ displacement, the calculations provide the only significant difference with experiment, as both functionals predict outward shifts which are too small compared with experimental values. However, our result is roughly consistent with previous DFT calculations. For example, Hameeuw et al. [30] calculated an outward relaxation of only 0.06 Å, whereas the study by Kiejna et al. [33] reported a zero shift for this layer. The theoretical-experiment difference has been attributed by Sushko et al. to be due to the presence of soft anharmonic surface vibrational modes, which are not taken into account in the *ab initio* calculations [10].

**Table 1. Atomic displacements (Å) along the (110) direction calculated for the relaxed four layer slab model compared with the experimental data. The atoms refer to those shown in the top layer in figure 1.**

| | **This work** | | **Experiment** | |
|---|---|---|---|---|
| **Atom** | PBE | optB88-vdW | LEED[a] | SXRD[b] |
| **$Ti_{6f}$** | 0.20 | 0.17 | 0.25 ± 0.03 | 0.25 ± 0.01 |
| **$Ti_{5f}$** | -0.11 | -0.15 | -0.19 ± 0.05 | -0.11 ± 0.01 |
| **O in-plane** | 0.19 | 0.14 | 0.27 ± 0.08 | 0.17 ± 0.03 |
| **$O_b$** | 0.02 | 0.01 | 0.10 ± 0.05 | 0.10 ± 0.04 |

**[a] Ref [46]. [b] Ref. [37].**

*3.2 Adsorption of methane*

We now consider the interactions of methane ($CH_4$) with the (110) surface. We used four starting configurations for methane, with one, two, or three CH bonds pointing towards the surface. In addition, two CH orientations were used; one with the plane of the CH bonds in parallel with the $Ti_{5f}$ row and the other rotated at 90$^o$, with the CH bonds perpendicular to the $Ti_{5f}$ row. Regardless of the initial configuration and the employed functional, the relaxed adsorbate always has two CH bonds oriented towards the surface perpendicular to the $Ti_{5f}$ row. Other configurations are not minima in the potential energy landscapes. Table 2 summarises the equilibrium distances at the relaxed adsorption configurations.

Two local minima were found for both functionals: one on top of the $Ti_{5f}$ cations and one on top of the $O_b$ sites. The stable adsorption configurations are shown in figure 2, together with the adsorption energies and geometric details.

In the PBE calculations it is found that the interaction between the surface and methane adsorbed on the $O_b$ site was very weak. Although this configuration corresponds to a shallow minimum of the potential energy, it gives a slightly positive energy with respect to the desorbed reference level, but the value is within the expected errors of the calculations so the sign here is meaningless. A weak adsorption was also found at the $Ti_{5f}$ site, with adsorption energy of -85 meV. In both cases, the configuration adopted by the methane molecule exhibits two CH bonds pointing towards the surface, and the plane containing these two bonds is perpendicular to the row of bridging oxygen atoms.

Using the optB88-vdW, the adsorption configurations are the same as in the PBE calculation, but the interactions become stronger. The adsorption energy at the $O_b$ site is now slightly negative, while adsorption at the Ti site becomes stronger by 0.27 eV. We have not found experimental information on this system to benchmark our calculations against, probably because such experiments would need to be done at very low temperature, considering the weak adsorption behaviour. However, our results clearly illustrate that taking into account the vdW interactions makes a significant difference in the description of the weak adsorption of methane on this surface. Similarly weak adsorption energies have been recently calculated for the adsorption of methane on Ni surfaces using a related vdW-enabled functional [47].

**Table 2. Interatomic distances calculated for the different adsorption modes of methane.**

| **Methane adsorption mode** | **Distance (Å)** | **PBE** | **optB88-vdW** |
|---|---|---|---|
| **$CH_4$-$O_b$** | C-$O_b$ | 3.47 | 3.24 |
| **$CH_4$-Ti** | C-Ti | 3.12 | 2.92 |

### *3.3 Adsorption of methanol*

For methanol ($CH_3OH$) we considered several initial orientations in the molecular state and one dissociated ($CH_3O^-$ + $H^+$) state. The starting configurations for molecular adsorption were with the $CH_3$ group or OH group orientated towards either the in-plane $Ti_{5f}$ or the bridging $O_b$ sites. The dissociated adsorption mode was calculated with the oxygen in the $CH_3O^-$ group interacting directly with the $Ti_{5f}$ site and the proton transferred to a neighbouring $O_b$ site.

In the least stable of the molecular adsorption configurations found here, $CH_3$-$O_b$ (Fig. 3a,b), the C-O bond in the molecule remains perpendicular to the surface, and the closest interaction is between the $CH_3$ group of methanol and the surface $O_b$ site. This adsorption mode is very weak (-0.05 eV) when calculated with PBE but it is stronger (-0.16 eV) when calculated using the optB88-vdW functional. Van der Waals interactions are responsible for a significant shortening (0.25 Å) of the distance between the surface and the physisorbed molecule.

The $CH_3$-$Ti_{5f}$ adsorption mode (Fig. 3c,d), where the $CH_3$ group is directly on top of the $Ti_{5f}$ site, is also relatively weak, but it is stabilised by a tilting of the molecule to form an OH…$O_b$ hydrogen bond. This adsorption configuration is stronger by 0.48 eV when using the optB88-vdW, for which the C-O bond in the molecule adopts an almost parallel orientation with respect to the surface, resulting in a much shorter OH…$O_b$, compared to the PBE geometry.

The configuration OH-Ti (Fig. 3e,f), with close interaction between the oxygen of the OH group and the $Ti_{5f}$, gave the most negative adsorption energy among the non-dissociated configurations. This adsorption mode is additionally stabilized by the non-bonding interaction between the proton in the hydroxyl group of methanol and a bridging oxygen at the surface.

However, the dissociated state, $CH_3O$-Ti (Fig. 3g, h), where the methanol oxygen atom also binds to the $Ti_{5f}$ site but the proton from the OH group is transferred to the bridging oxygen at the surface, has a more negative adsorption energy compared to the molecular state, both from PBE and optB88-vdW calculations. Interestingly, the energy difference between the most stable molecular adsorption mode and the dissociative adsorption mode is very similar for both functionals: 0.16 eV for PBE and 0.17 eV for optB88-vdW. These energy differences are also close to the value of 0.19 eV reported in early studies by Bates et al. [8,48] from GGA calculations with the PW91 functional [49]. Therefore the prediction of favourable dissociation is not affected by the consideration of van der Waals interactions.

It is interesting to note here that more recent GGA-PW91 calculations from Sánchez de Armas et al. [9] found that the molecular and dissociated states of methanol adsorption are almost degenerate on the $TiO_2$ (110) surface. They actually found that the energy difference depend on the thickness of the symmetric slab model used to represent the surface, and the difference in the average from five- and six-layer slabs was only ~0.01 eV (on the stoichiometric surface). As we discussed in the methodology section, by fixing the bottom layers of our slab we can expect our results to converge faster with the slab thickness than in symmetric slab calculations. In order to check this convergence, we have performed adsorption energy calculations in a thicker five-layer slab (again with the two bottom layers fixed to bulk positions) for the most stable molecular and dissociated configurations, with both PBE and optB88-vdW, to compare with our four-layer slab results. We find that none of the calculated adsorption energies changes by more than 0.03 eV. The difference between molecular and dissociative adsorption energies changes by less than 0.02 eV with either functional, and the differences between PBE and optB88-vdW adsorption energies change by less than 0.01 eV when the thickness of the slab is increased. Therefore we can conclude that the prediction of favourable dissociation is robust not only with respect to the functional employed (including or excluding vdW interactions) but also with respect to the thickness of the simulation slab.

Table 3 shows the equilibrium distances for the relaxed adsorption configurations. In the most stable molecular and dissociative adsorption modes the molecule has a similar orientation with respect to the surface, as also noted by Sánchez de Armas et al. [9]. However, in the case of the dissociated configuration, a hydrogen bond is formed between the surface OH group and the oxygen in the $CH_3O^-$ fragment, involving a significant buckling of the surface. An upward displacement of 0.33 Å for the optB88-vdW and 0.40 Å for PBE occurs in the $Ti_{5f}$ atom in order to facilitate the formation of the hydrogen bond.

Another interesting aspect about the adsorption of methanol at the $TiO_2(110)$ surface is whether an activation barrier exists for the transfer of the hydroxyl proton to the surface. The GGA-PW91 study of Bates et al. [48] concluded, from nudged elastic band (NEB) calculations [50], that the proton transfer was barrierless, while the molecular dynamics simulations of Sánchez de Armas et al. [9] indicated that any transition barrier would be small. We have performed NEB calculations with the two functionals being compared in our study; the energy profiles are shown in Fig. 4. In the case of PBE calculations, the highest energy found in the dissociation path is only a few meV above the molecular adsorption minimum. Such a small energy difference is actually not meaningful within the precision of our calculations (although it was enough to prevent the spontaneous transfer of the proton during the local optimisation of the molecular adsorption state); hence we conclude that the transition is essentially barrierless at PBE level. On the other hand, in the energy profile calculated with the optB88-vdW functional, a still small but clear activation barrier of ~0.10 eV is present.

The comparison of our results with experimental findings is made difficult by conflicting reports in the literature about this system. Methanol adsorption at the $TiO_2$(110) surface has been suggested in some studies to be predominantly molecular (e.g. [51,52]) while other experiments have indicated that dissociation takes place at the surface (e.g. [53-55]). The temperature programmed desorption (TPD) study by Henderson et al. [52] reported the presence of both molecular and dissociated methanol at the $TiO_2$(110). They found three desorption peaks (at 295 K, 350 K and 480 K) for methanol on the vacuum-annealed surface of $TiO_2$ (110). The 295 K peak was the most prominent and was assigned to the molecular adsorption configuration, while the higher temperature peaks were assigned to two dissociative configurations: the 350 K peak corresponds to methoxyls at non-vacancy sites (as calculated here), and the one at 480 K to methoxyls at oxygen vacancy sites. The latter situation, which should correspond to the most negative adsorption energy, has not been considered here because we only study the (1×1) stoichiometric surface. This interpretation of the TPD experiment suggests that the dissociative adsorption of methanol is stronger than molecular adsorption, in qualitative agreement with our calculations. However, if we attempt to extract adsorption energies from these data using the Redhead method [56], assuming typical pre-exponential factors in the range $10^{13}s^{-1}$ – $10^{15}s^{-1}$, we get adsorption energy values between -0.80 and -0.91 eV for the 295 K peak and values between -0.95 eV and -1.09 eV for the 350 K peak. Our optB88-vdW calculations predict much more negative adsorption energies for the molecular and dissociated states, respectively. Our PBE results, although still stronger than expected, are much closer than the optB88-vdW results to the TPD/Redhead values. This disagreement requires further analysis. In their study of acetaldehyde adsorption at $TiO_2$(110), Plata el al. [25] concluded that the optB86b-vdW functional (from the same family of vdW-enabled functionals) overestimated the strength of adsorption at low coverage. They ascribed the overestimation to double counting of some short-range exchange-correlation terms in the functional, as also suggested by Göltl and Hafner in their assessment of the original vdW-DF functional for alkane adsorption on chabazite [57]. It is therefore possible that our optB88-vdW calculations are similarly overestimating the adsorption strength of chemisorbed molecules at the $TiO_2$ surface. However, it is important to note that the comparison with experimental values obtained via a Redhead analysis of the TPD experiments of Henderson et al. might also be problematic. For example, they found that the peak positions changed significantly depending on methanol exposure, which is a situation for which a Redhead analysis is generally not valid. Furthermore, our calculations correspond to an ideal stoichiometric and defect-free surface, while their TPD measurements were made on real surfaces including defects like oxygen vacancies and surface steps and corners, which complicates the interpretation. Clearly, this is a topic that requires further theoretical and experimental attention.

**Table 3. Interatomic distances calculated for the different adsorption modes of methanol.**

| Methanol adsorption mode | Distance (Å) | PBE | optB88-vdW |
|---|---|---|---|
| **$CH_3$-$O_b$** | C-$O_b$ | 3.27 | 3.02 |
| **$CH_3$-Ti** | C-Ti | 3.34 | 3.12 |
| | HO-$O_b$ | 3.66 | 1.87 |
| **OH-Ti** | O-Ti | 2.16 | 2.15 |
| | HO-Ob | 1.75 | 1.77 |
| **$CH_3$O-Ti** | O-Ti | 1.87 | 1.87 |
| | O-$H_{Ob}$ | 2.18 | 2.25 |

### *3.4 Adsorption of formic acid*

The next adsorbate considered in our study is formic acid (HCOOH). With oxygen atoms in the adsorbate, the dominant interactions are expected to be those between these oxygen atoms and the coordinatively unsaturated Ti atoms at the surface. Following Sushko et al. [10], we consider both monodentate and bidentate adsorption configurations of this molecule, where one or two oxygen atoms of the molecule interact closely with the surface. We also compare the stability of molecular versus dissociative adsorption, using the two density functionals.

In principle, for molecular adsorption there are two possible bidentate configurations, with either one or two surface Ti atoms participating in the interaction with the molecule ("bidentate1" and "bidentate2" respectively). However, in contrast with the work of Sushko et al. [10], where a shallow minimum for the bidentate1 configuration for the formic acid molecule was reported, we found here that for both functionals the bidentate1 adsorption configuration was not a minimum in the energy landscape and converged to a bidentate2 configuration upon relaxation. A very recent study using the PBE functional (with no dispersion correction) has also found only one bidentate mode for molecular adsorption [31]. Our PBE calculations predict the adsorption energy of the formic acid monodentate configuration to be almost exactly equivalent to the bidentate2 at the $TiO_2$ (110) surface. Sushko et al. predicted that the adsorption of formic acid in the bidentate2 configuration is ~0.1 eV lower in energy compared to the monodentate configuration [10]. The relaxation observed in our calculation of the monodentate molecular orientation leads to additional interaction between the carbon's hydrogen atom and the $O_b$ of the surface (Fig. 5a), which results in a monodentate adsorption energy similar to that in the bidentate2 configuration.

However, when the optB88-vdW is used to account for the dispersion forces, the adsorption in the monodentate and bidentate2 configurations of (non-dissociated) formic acid is more exothermic by approximately 0.5 eV and 0.6 eV, respectively, than when PBE is used. Thus, whereas the PBE adsorption energy values for monodentate and bidentate2 configurations are essentially the same, the optB88-vdW predicts adsorption in the bidentate2 configuration to be stronger (by 0.1 eV) than in the monodentate configuration.

Since there is a considerable body of experimental evidence showing that formic acid dissociates at the $TiO_2$(110) surface [31,58-62], we have also considered different modes of dissociative adsorption (Fig 5e-h). The acid dissociation occurs by the transference of the oxygen bound proton to a neighbouring $O_b$ site at the surface. The adsorption of the formate bidentate2 ion (Fig. 5g,h) was clearly favoured over the monodentate ion (Fig. 5e,f) using both PBE and the optB88-vdW. The bidentate2 ion is bound to the substrate $Ti_{5f}$ cations with the molecule orientated along the (001) direction. The O-C-O intramolecular bond angle was found to be 127.1$^o$ using the optB88-vdW which represents a widening from 122.2$^o$ in the monodentate ion due to the interaction with the two $Ti^{4+}$ surface ions. The value of the bound bidentate2 formate bond angle is within the experimental error of the experimental bond angle of 126 ± 4$^o$ as measured by Thevuthasan et al. using high-energy photoelectron diffraction [59]. These authors also reported a Ti(surface)-O(formate) vertical distance of 2.1 ± 0.1 Å. The DFT calculated data predicts an average vertical separation between the Ti lattice cations and the formate oxygens of 2.09 for PBE and 2.08 Å for the optB88-vdW, both in reasonable agreement with experiment.

**Table 4. Interatomic distances calculated for the different adsorption modes of formic acid.**

| Formic acid adsorption mode | Distance (Å) | PBE | optB88-vdW |
|---|---|---|---|
| **HCOOH monodentate** | $O_{CO}$-Ti | 2.19 | 2.16 |
| | HC-$O_b$ | 2.49 | 2.35 |
| **$HCOO^-$ monodentate** | $O_{CO}$-Ti | 1.92 | 1.92 |
| | HC-$O_b$ | 2.65 | 2.60 |
| **HCOOH bidentate2** | $OH_{HCOO}$-Ti | 2.43 | 2.43 |
| | $O_{HCOO}$-Ti | 2.27 | 2.25 |
| **$HCOO^-$ bidentate2** | $O_{HCOO^-}$-Ti | 2.10 | 2.09 |
| | $O_{HCOO^-}$-Ti | 2.08 | 2.07 |

As expected from experiment, dissociative adsorption of formic acid is much more stable than molecular adsorption. For the most stable adsorption configuration (bidentate2), the energy lowering from dissociation at the surface was 1.30 eV using PBE and 1.37 eV using optB88-vdW. This energy difference is much larger than the corresponding value found for methanol dissociation. The stronger tendency towards deprotonation of the carboxylic group adsorbed in the symmetric bidentate mode can be explained by resonance: the symmetry between the two O atoms in the binding $COO^-$ group leads to a stabilizing delocalization of the negative charge.  Not surprisingly, this bridging mode has been reported to be the most common configuration for species containing a carboxyl group, when adsorbed at $TiO_2$ surfaces [7,63,64].

### *3.5 Adsorption of glycine*

Glycine, the smallest of the amino acids, constitutes a simple model of biomolecule including both carboxylic and amino groups. For glycine adsorption on $TiO_2$(110) we used eight initial configurations based on the neutral ($H_2N$-$CH_2$-COOH), zwitterionic ($H_3N$-$CH_2$-COO) and anionic ($H_2N$-$CH_2$-$COO^-$ + $H^+$) forms of adsorption found in a recent DFT study (without dispersion corrections) by Tonner  [11]. Our notation for the configurations follows the ones used in that study: N=neutral, ZW=zwitterionic, and AN=anionic. In brackets we give information about the binding configuration, e.g. AN(OO) means anionic glycine bonded to the surface via two oxygen atoms. An H subscript indicates the presence of a hydrogen bond to the surface. The meaning of each label is obvious from Figures 6 and 7.

Glycine shows strong binding to the $Ti_{5f}$ ion either via the amine $NH_2$ group or the carboxyl group. The strongest adsorption (as calculated by either functional) occurs for the bidentate

AN and ZW forms via the carboxyl group. These bidentate modes, where the two oxygen atoms in the carboxylic group attach directly to the 5-fold-coordinated Ti sites at the surface, are analogous to the most favourable adsorption mode of formic acid described above. However, unlike carboxylic acids, glycine can also form hydrogen bonds with the surface via the amino group, thus providing additional stability, and leading to a tilting of the molecule with respect to the "vertical" configuration. Comparison of the PBE adsorption energy for the AN(OO) configuration ($E_{ads}$ = -2.24 eV) with that of the most stable mode, $AN_H$(OO) ($E_{ads}$ = -2.41 eV), provides a value of 0.17 eV for the stabilisation energy associated to the tilting. DFT calculations by Tonner resulted in a similar tilting stabilisation effect of the surface bound hydrogen atom and the nitrogen atom of the amino group [11]. It is interesting that in the case of calculations including vdW interactions, the "vertical" configurations AN(OO) and ZW(OO) are not minima in the adsorption landscape. This suggests that vdW interactions stabilise the tilted configurations even further, by reinforcing the attraction between the amino group and the surface. The expansion of the basin of attraction of the tilted configuration minimum is probably responsible for the disappearance of the weaker minimum at the "vertical" configuration.

Calculations with either functional show the bidentate $ZW_H$(OO) mode to be higher in energy (by 0.11 eV for PBE and by 0.10 eV for opt-B88vdW) than the corresponding anionic mode ($AN_H$(OO)), i.e. dissociation of the glycine molecule at the surface is predicted to be favourable. The $ZW_H$(OO) configuration was calculated to be the most stable mode in a previous DFT study by Ojamäe et al. and co-workers [65], which ignored dissociated configurations. However, the lower stability of the zwitterionic mode compared with the anionic mode in our calculations is in agreement with the scanning tunnelling microscopy (STM) experiments by Barteau and Qiu, who found no evidence for the zwitterionic structure on the $TiO_2$ surface [66]. Our prediction is also in agreement with the photoelectron spectroscopy experiments by Soria et al., who found that dissociative adsorption is dominant at low coverage, while only at high (multi-layer) coverage the zwitterionic mode becomes dominant [67]. Photoelectron diffraction (PhD) experiments by Lerotholi et al. [68] also found that at low coverage only the deprotonated glycinate species is present, and a coadsorbed bridging OH species was also identified, confirming the occurrence of deprotonation. It is worth noting that, although our calculated value for the energy difference between the $AN_H$(OO) and $ZW_H$(OO) modes seems small, it is large enough to explain the dominance of dissociative adsorption. A simple Boltzmann's statistical analysis shows that in equilibrium at 300 K, an energy difference of 0.10 eV in favour of dissociation leads to 98% dissociative adsorption.

We have also calculated the activation barriers for the glycine deprotonation reaction at the surface, i.e., the transition from the $ZW_H$(OO) to the $AN_H$(OO) configurations, using the nudged elastic band method [50]. We find that vdW interactions have very little effect on the transition barrier: the activation energies are 0.48 eV for PBE and 0.45 eV for optB88-vdW. This barrier is not high enough to prevent the dissociation of glycine adsorbed at room temperature, which is consistent with the experimental observation of predominantly dissociative adsorption at low coverage.

The main effect of the inclusion of vdW effects in the calculation is therefore the prediction of stronger adsorption energies in all configurations, together with the disappearance of some of the modes observed in the PBE calculation (Figures 6 and 7). The absence of the 'vertical' modes AN(OO) and ZW(OO) was already discussed above. In addition, the monodentate zwitterionic mode $ZW_H$(O), which is weak but still stable in the PBE calculation, is also absent when vdW interactions are taken into account.

A point that remains to be clarified experimentally is the prediction of tilting of the glycine adsorbate at the surface. Neither the STM study by Qiu and Barteau [66] nor the PhD study by Lerotholi et al. found any evidence of strong interaction of the amino group with the surface. If the interaction of the amino group is sufficiently strong it should prevent free rotation about the C-C bond giving rise to asymmetric features in STM, which were not seen. In the PhD experiment, the conclusion that the amino N atom is not involved in the surface bonding is supported by the absence of significant modulations of the N 1s emission. Therefore, Lerotholi et al. also suggested that the molecule is in a "vertical" or "standing-up" position. In contrast, our calculations strongly suggest that the tilting to form hydrogen bond via the amino group is very favourable, and the vertical mode of adsorption is not even metastable when vdW interactions are included in the simulation. As pointed out by Qiu and Barteau, an STM experiment at lower temperatures and with higher resolution would be required to provide a more definite answer to this question.

Finally, it is interesting to compare glycine adsorption at this oxide surface with the adsorption at metal surfaces, which typically have higher densities of adsorption sites of a given type. For example, when glycine adsorbs on Cu surfaces, it binds both via the carboxylic oxygen atoms and via the amino group, adopting a fully "horizontal" configuration [69,70]. In the case of $TiO_2$, two neighbouring $Ti_{5f}$ sites are available for the adsorption of the carboxylic group, but there are no other $Ti_{5f}$ sites at the required distance within the surface for simultaneous binding of glycine via the amino group. Therefore a "horizontal" adsorption mode of glycine at $TiO_2$(110) is not likely, as has been pointed out by Lerotholi et al. [68]. Our results show that the preferred mode of adsorption is not fully "vertical" either, but tilted, in order to favour the formation of hydrogen bonds between the surface and the amino group.

**Table 5. Interatomic distances calculated for the different adsorption modes of glycine.**

| **Glycine Adsorption mode** | **Distance (Å)** | **PBE** | **optB88-vdW** |
|---|---|---|---|
| $AN_H$(OO) | $O_{COOH}$-Ti | 2.08 | 2.08 |
| | $O_{COOH}$-Ti | 2.09 | 2.07 |
| | $O_b$-H | 1.68 | 1.67 |
| AN(OO) | $O_{COOH}$-Ti | 2.06 | - |
| | $O_{COOH}$-Ti | 2.07 | - |
| $AN_H$(O) | O-Ti | 1.96 | 1.95 |
| | $O_b$-H | 1.47 | 1.48 |
| $ZW_H$(OO) | $O_{COOH}$-Ti | 2.12 | 2.12 |
| | $O_{COOH}$-Ti | 2.12 | 2.12 |
| | $H_N$-$O_b$ | 1.66 | 1.67 |
| | $H_N$-$O_b$ | 1.67 | 1.67 |
| ZW(OO) | $O_{COOH}$-Ti | 2.18 | - |
| | $O_{COOH}$-Ti | 2.21 | - |
| $ZW_H$(O) | $O_{COOH}$-Ti | 2.02 | - |
| | $H_N$-$O_b$ | 1.75 | - |
| AN(ON) | $O_{COOH}$-Ti | 1.91 | 1.91 |
| | N-Ti | 2.26 | 2.24 |
| N(N) | N-Ti | 2.30 | 2.27 |

## 3. Conclusions

We have presented a DFT investigation of the adsorption of small organic molecules at the $TiO_2$(110) surface, contrasting two different density functionals: the generalized gradient approximation functional PBE, and the non-local correlation functional optB88-vdW, which is capable of correctly describing vdW interactions. Our results show that the inclusion of vdW effects can produce some significant differences in the adsorption geometries of the molecules. As one might expect, the effect of vdW interactions on the adsorption geometry is more important when there are no other strong bonding interactions. To illustrate this point, the variations introduced by optB88-vdW (compared to PBE) in the equilibrium distances between the molecular hydrogen atoms and surface oxygen atoms, are plotted in Fig. 8 against the adsorption energies (from PBE). The cases of stronger adsorption mediated by carboxylic groups are excluded from the plot. This analysis clearly shows that as the adsorption strength decreases, the effect of the vdW interactions on the adsorption distances becomes stronger.

Methane adsorption was found to be much stronger when using optB88-vdW (-0.36 eV) than when using PBE (-0.09 eV) at the most stable configuration. This is as expected, considering that methane is a non-polar molecule for which other interactions, besides dispersion, are very weak.

However, although hydroxyl, carboxyl and amino groups all possess a permanent molecular dipole moment, they also showed an increase in the adsorption strength when vdW interactions where included. Methanol adsorption energies were 0.1-0.5 eV (depending on configuration) stronger when using optB88-vdW. Both functionals predict that the dissociated state of methanol is more favourable at the surface than the molecular state, with the energy difference between the two states being very similar for both functionals. Furthermore, the prediction of favourable dissociation is robust with respect to the thickness of the simulation slabs, in contrast with previous reports. But while the dissociation of methanol is predicted to be essentially barrierless with PBE, the optB88-vdW functioned predicted a small barrier of ~0.10 eV.

The adsorption of formic acid was found to be between 0.45 eV and 0.65 eV (depending on configuration and dissociation) stronger with the optB88-vdW functional. However, the nature of the preferred adsorption configuration of formic acid is also the same for both functionals. The carboxyl group is likely to attach to this surface by binding with each O atom occupying a single $Ti_{5f}$ adsorption site, and dissociation is always clearly favourable over molecular adsorption.

The vdW forces can also affect the number of stable adsorption modes of larger organic molecules with respect to the surface, as seen with the molecular orientations of glycine obtained using optB88-vdW. For example, when vdW forces are considered, the “vertical” mode of adsorption, where the amino group does not interact with the surface, is not stable. Both the PBE and the optB88-vdW functionals predict that the anionic mode, where a proton is transferred to form a surface hydroxyl group, is more stable than the non-dissociated zwitterionic mode. Like for the other adsorbates, the strength of glycine adsorption is significantly higher when vdW effects are considered, by 0.2-0.6 eV.

The number of adsorbates studied in our paper is too small to provide a general quantitative rule for estimating the adsorption energies for larger classes of molecules. However, our results, together with previous studies, make it possible to infer some qualitative conclusions applicable to other molecules with similar structural motifs. For example, the most stable binding mode for both formic acid and glycine, which is the bidentate adsorption via the carboxylic group oxygen atoms to two $Ti_{5f}$ sites, involving the transfer of a proton to the surface forming a hydroxyl group, is likely to apply to other larger molecules containing the COOH group. The observation that methanol also dissociates suggests that the availability of both unsaturated Ti sites and bridging O sites at the $TiO_2$(110) surface generally provides a strong driving force towards the deprotonation of hydroxyl groups in adsorbates. Our results show that these conclusions are actually not affected by the inclusion of van der Waals effects in the simulations.

As a general comment, PBE and optB88-vdW differ significantly in the prediction of the adsorption geometries and energies of weakly adsorbed (physisorbed) molecules. In the case of strongly adsorbed (chemisorbed) molecules both functionals give similar results for the adsorption geometry, and also similar values of the relative energies between different adsorption modes of a given molecule. However, the absolute adsorption energy values are significantly different between the two functionals in all cases, with optB88-vdW always predicting stronger adsorption than PBE. While the comparison with experiment is difficult due to the complexity of experimental conditions, the analysis of the methanol case suggests that the optB88-vdW might be overestimating the strength of surface-adsorbate interactions in chemisorbed configurations.

## Acknowledgments

Via our membership of the UK's HPC Materials Chemistry Consortium, which is funded by EPSRC (EP/L000202), this work made use of HECToR and ARCHER, the UK's national high-performance computing services. MT is grateful to UCL's Industrial Doctorate Centre on Molecular Modelling & Materials Science, and to Straumann AG for the sponsorship of his PhD.

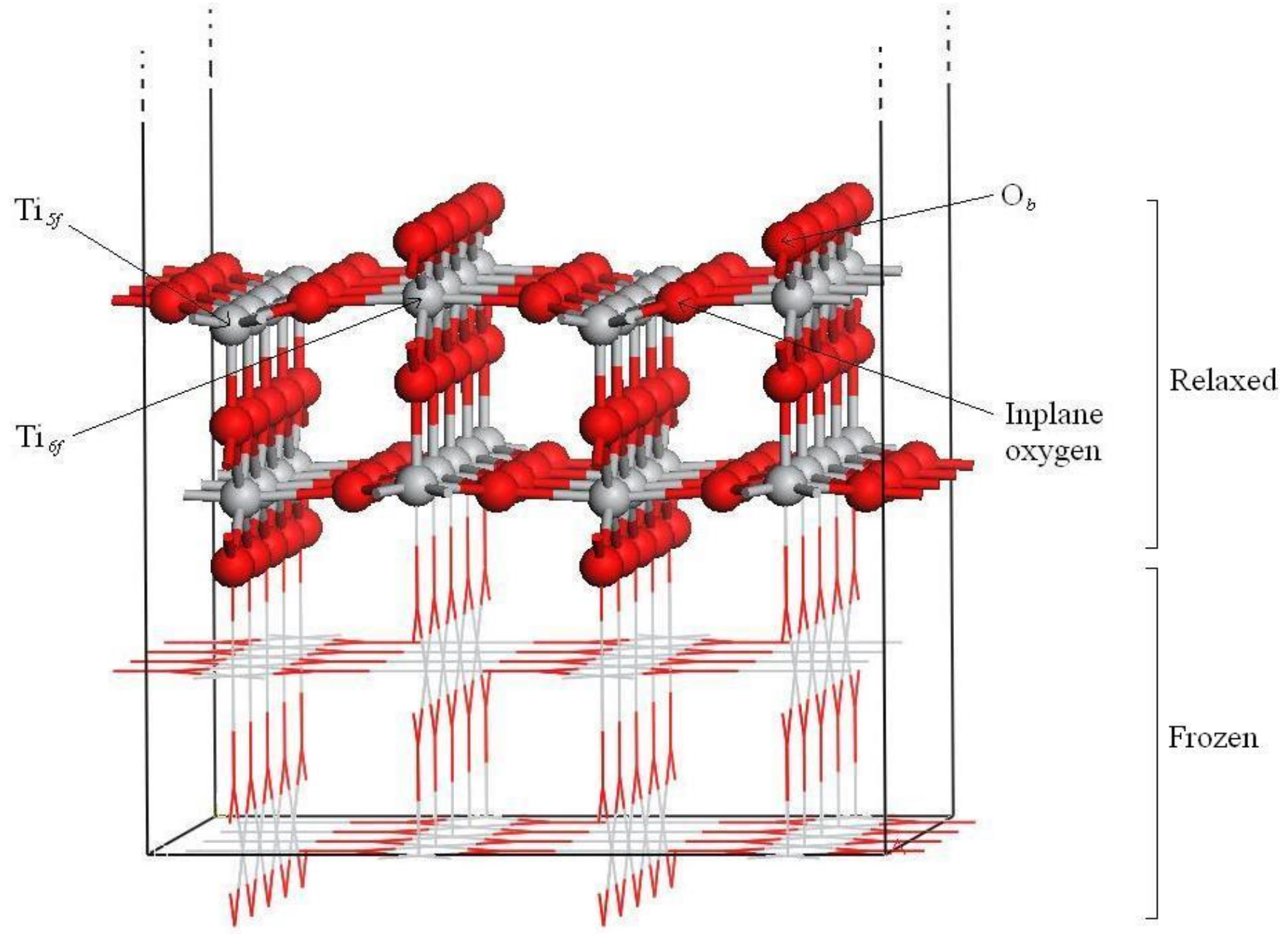

**Fig. 1. $Ti_{64}O_{128}$ slab cell used in our calculations to represent the rutile (110) surface. Oxygen atoms are in red, titanium atoms are in grey.**

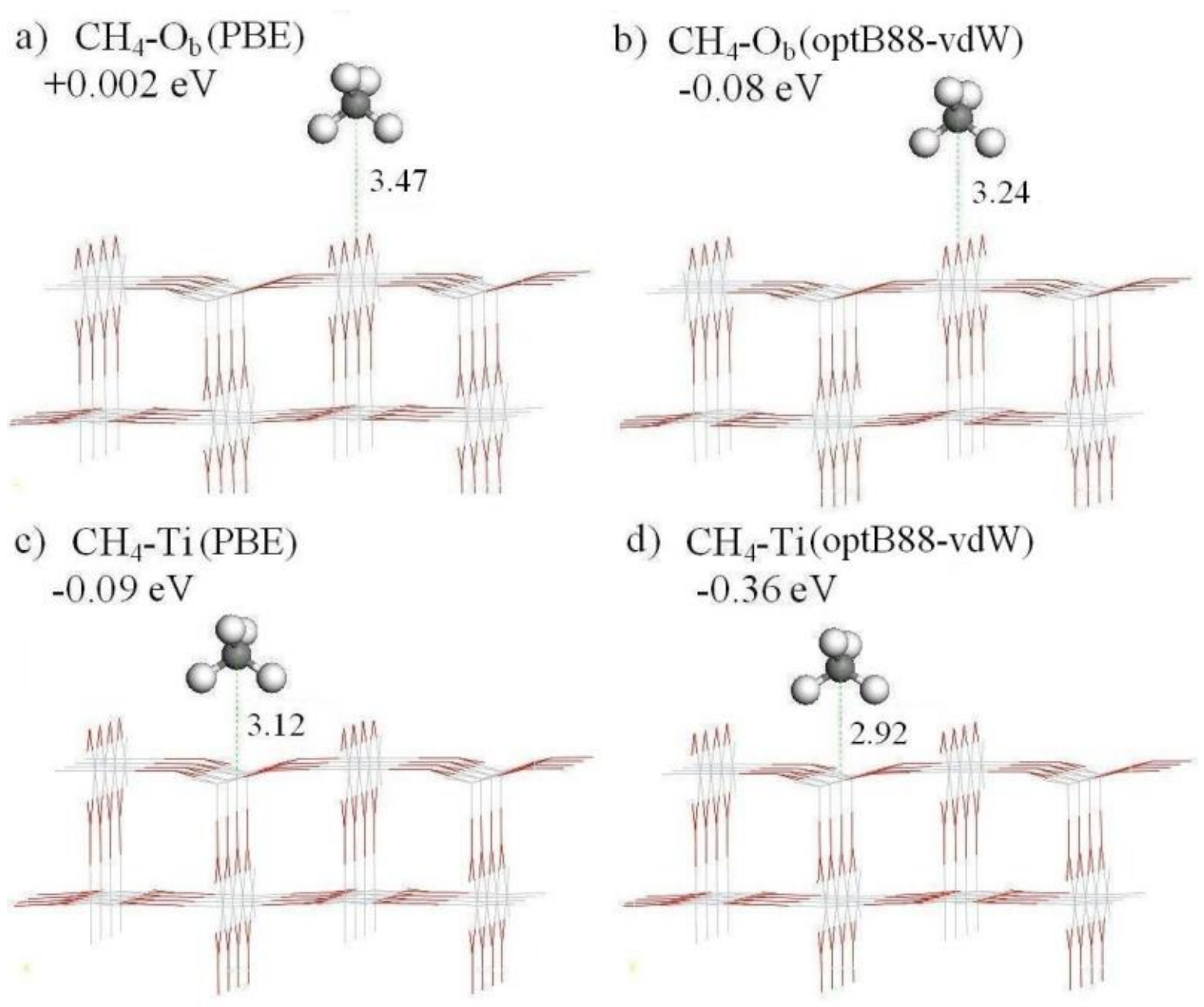

**Fig. 2. Adsorption modes of methane at the bridging oxygen site ($CH_4$-$O_b$) and at the five-fold-coordinated Ti sites ($CH_4$-Ti). The values given in eV are the adsorption energies $E_{ads}$. Some interatomic distances are given (in Å), next to the dotted lines joining the two atoms.**

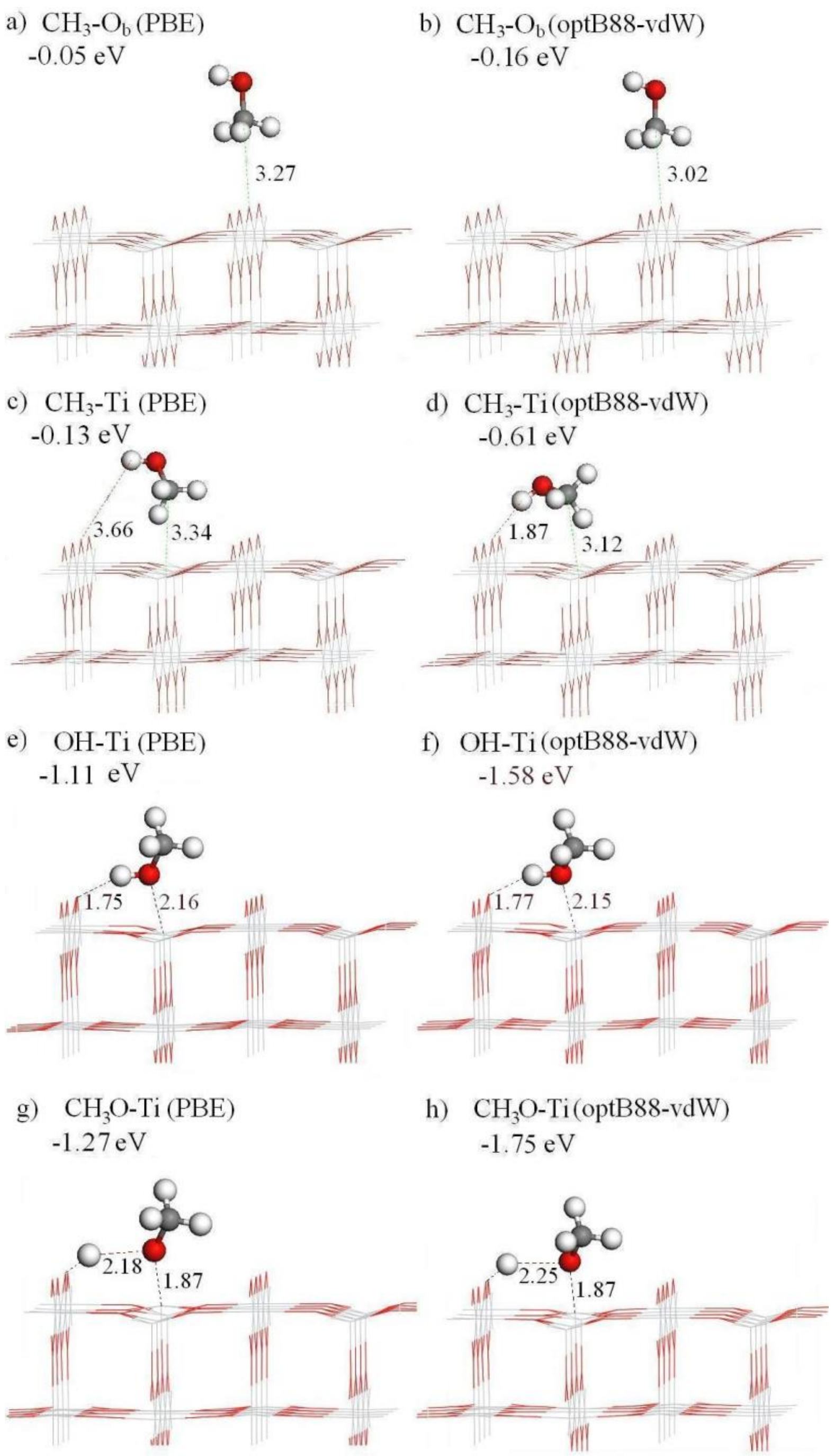

**Fig. 3. Molecular (a-f) and dissociative (g-h) adsorption modes of methanol.**

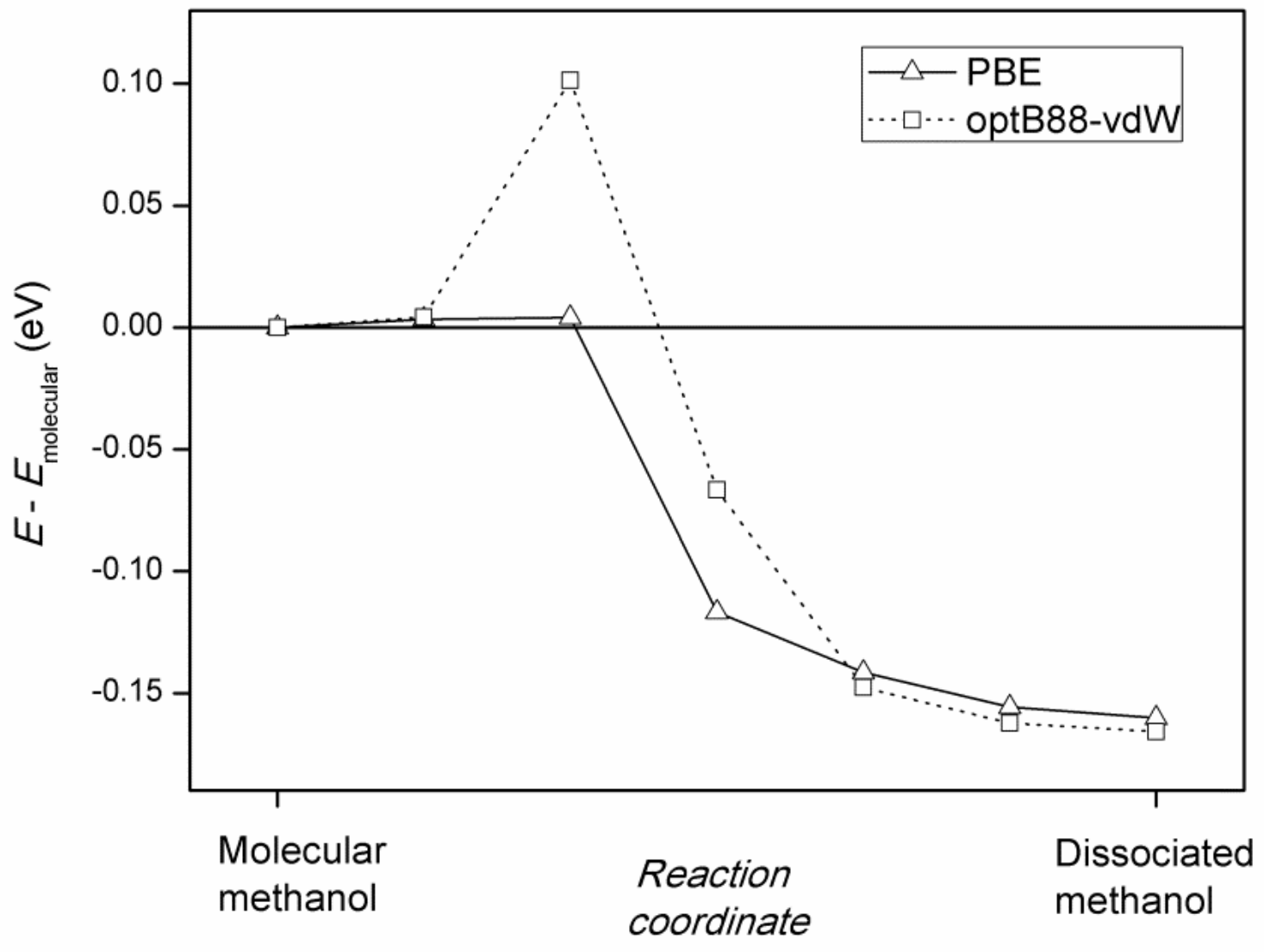

**Fig. 4. Nudged elastic band energy profiles for methanol dissociation at the $TiO_2$(110) surface, as calculated with the PBE and optB88-vdW functionals. The zero energy is set at the molecular adsorption state for each functional.**

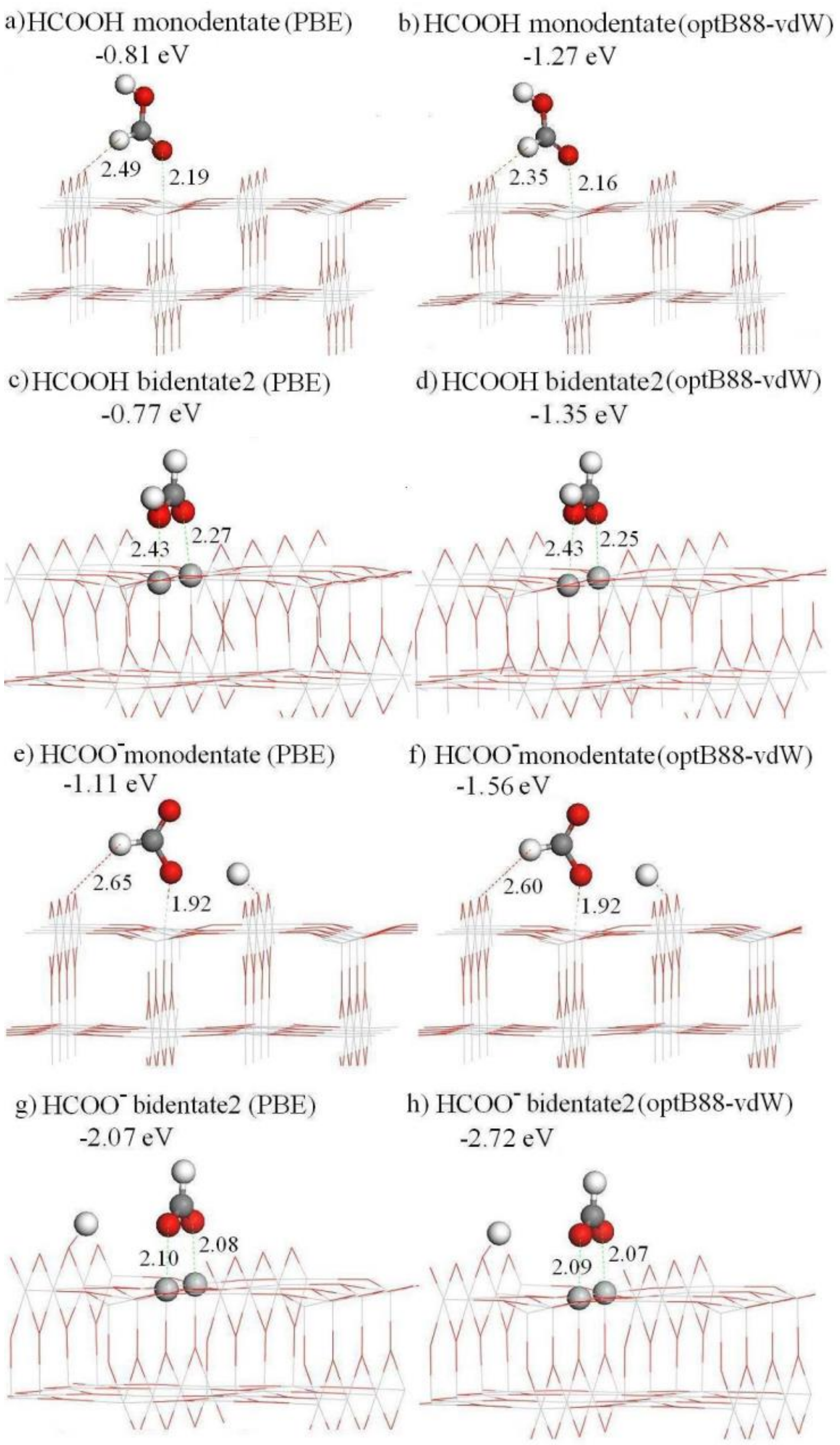

**Fig. 5. Molecular (a-d) and dissociative (e-h) adsorption modes of formic acid.**

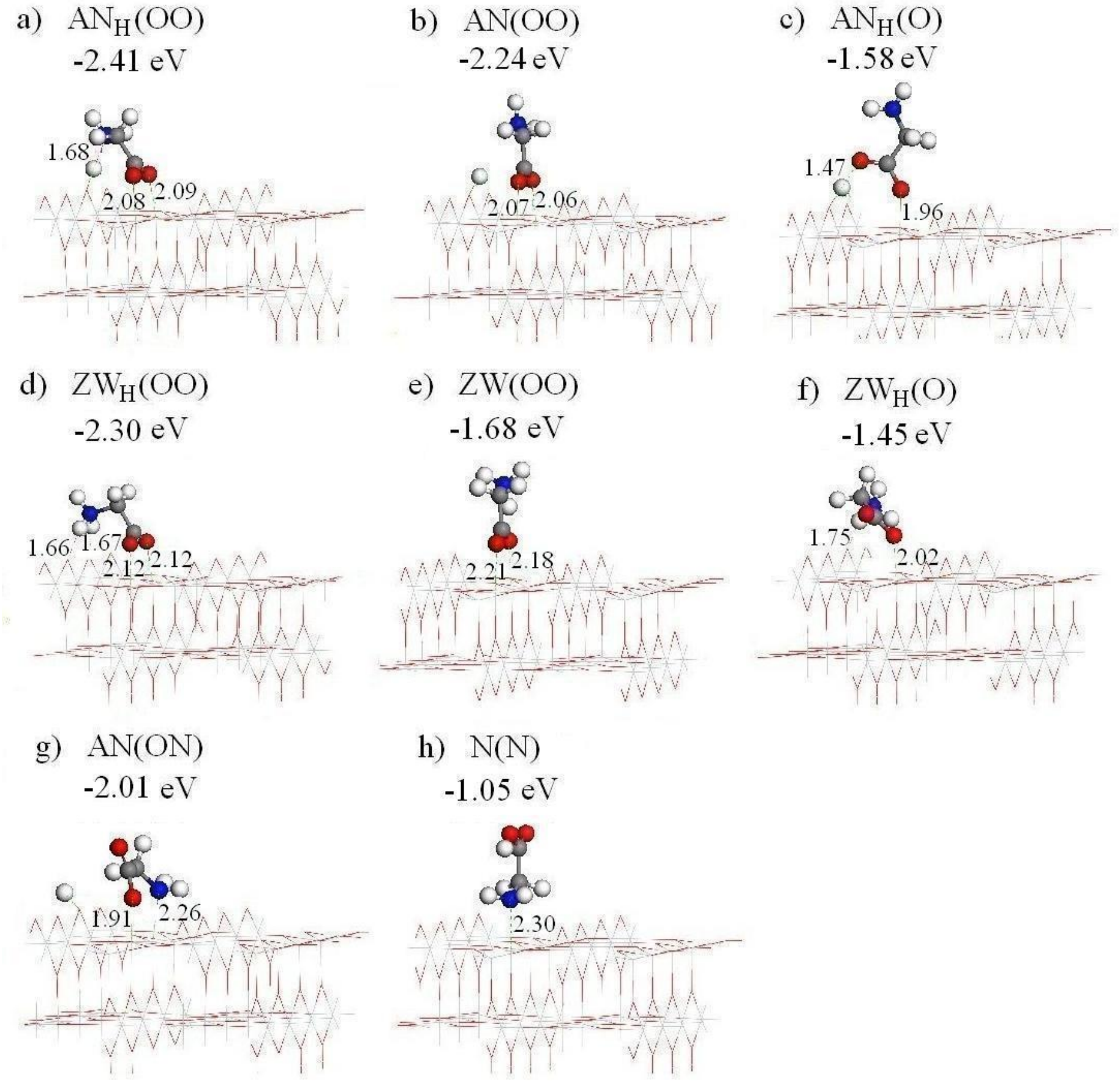

**Fig. 6. Adsorption modes predicted by the PBE calculations for anionic (AN), zwitterionic (ZW) and neutral (N) forms of glycine. Anionic glycine is formed from the transfer of the zwitterionic $NH_3^+$ proton to the $O_b$ to form a hydroxyl group at the surface.**

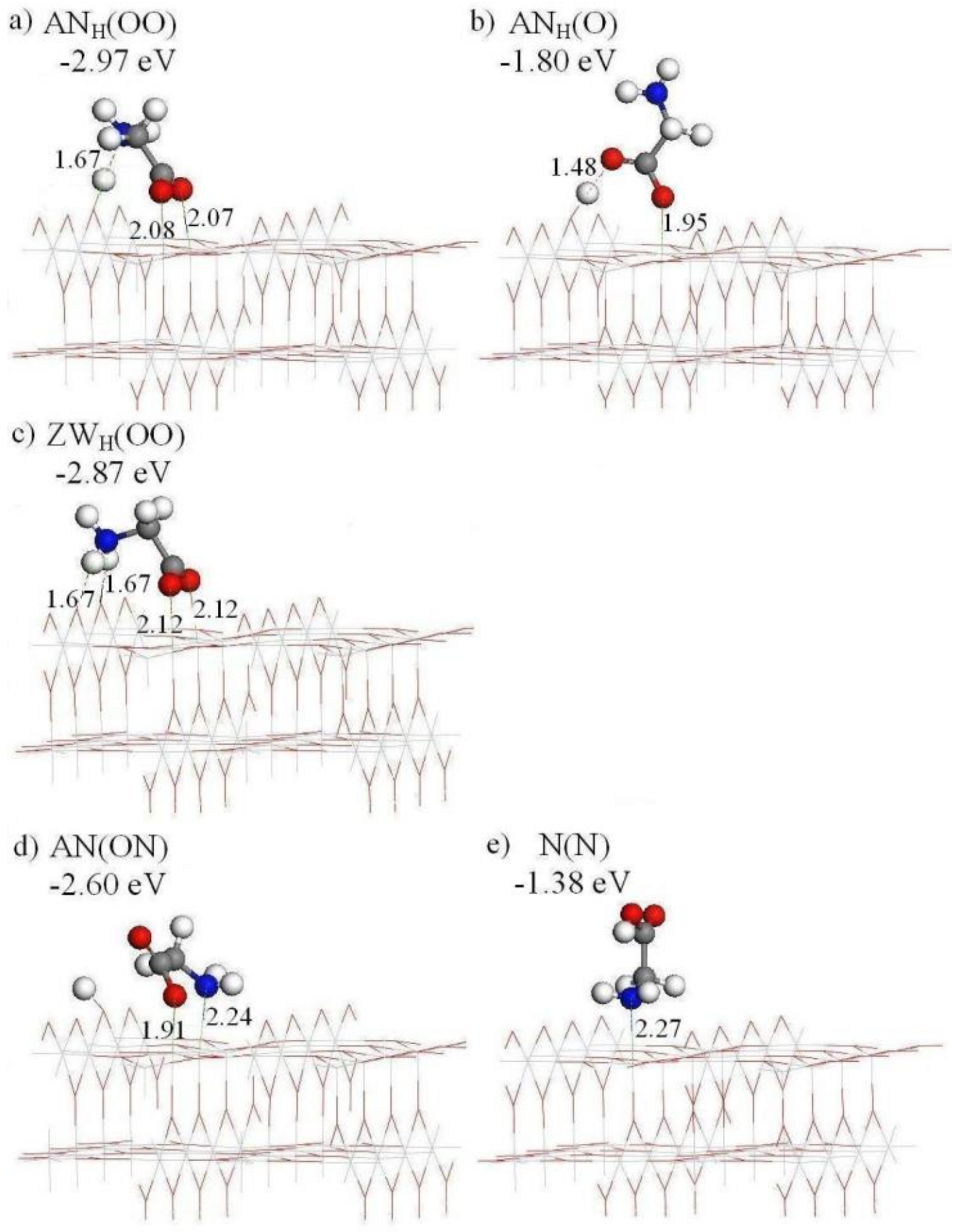

**Fig. 7. Adsorption modes predicted by the optB88-vdW calculations for anionic (AN), zwitterionic (ZW) and neutral (N) forms of glycine.**

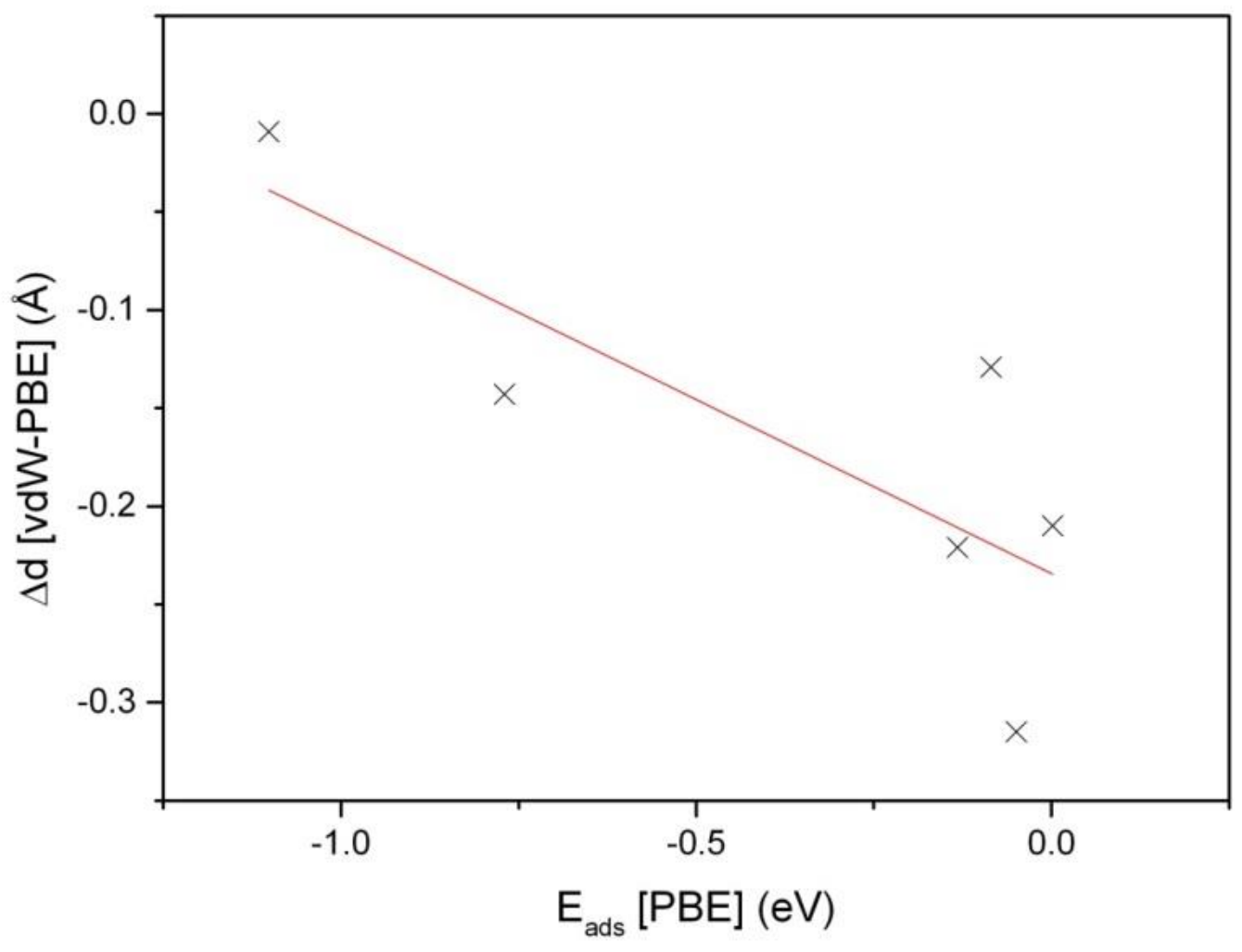

**Fig. 8. Differences in the $H_{(molecule)}$-$O_{b(surface)}$ distances calculated by optB88-vdW and the PBE functionals, plotted against the PBE adsorption energy. Results are included only for formic acid, methane and methanol. The straight line is only a guide to the eye.**